# Structure and magnetic properties of (Fe$_2$O$_3$)$_n$ clusters ($n$ = 1-5)

A. Erlebach, C. Hühn, R. Jana, and M. Sierka[*]

*Otto-Schott-Institut für Materialforschung, Friedrich-Schiller-Universität Jena,*

*Löbdergraben 32, 07743 Jena, Germany*

Global minimum structures of neutral (Fe$_2$O$_3$)$_n$ clusters with $n$ = 1-5 were determined employing genetic algorithm in combination with *ab initio* parameterized interatomic potentials and subsequent refinement at the density functional theory level. Systematic investigations of magnetic configurations of the clusters using a broken symmetry approach reveal antiferromagnetic and ferrimagnetic ground states. Whereas (Fe$_2$O$_3$)$_n$ clusters with $n$ = 2-5 contain exclusively Fe$^{3+}$, Fe$_2$O$_3$ was found to be a special case formally containing both Fe$^{2+}$ and Fe$^{3+}$. Calculated magnetic coupling constants revealed predominantly strong antiferromagnetic interactions, which exceed bulk values found in hematite. The precise magnetization (spin) state of the clusters has only small influence on their geometric structure. Starting from $n$ = 4 also the relative energies of different cluster isomers are only weakly influenced by their magnetic configuration. These findings are important for simulations of larger (Fe$_2$O$_3$)$_n$ clusters and nanoparticles.

[*] E-mail: marek.sierka@uni-jena.de

Nanoparticulate (NP) iron oxides in form of clusters and nanoparticles find a number of applications owing to their unique magnetic, biochemical and catalytic properties.[1] These include applications in catalysis, biomedical uses like magnetic hyperthermia in cancer treatment, targeted drug release, magnetic resonance imaging and immunoassays as well as magnetic data storage.[2-5] Convenient synthesis methods for such nanoparticles are gas phase processes, such as flame spray pyrolysis,[6] laser ablation,[7] plasma synthesis[8] and the laser vaporization of iron oxide raw powders.[9] Small clusters and particles are important intermediates in early stages of the nanoparticle formation processes and the knowledge of their atomic structure and properties are crucial components for establishing detailed nucleation and growth mechanisms.[10] A greater understanding of these mechanisms is essential to control and optimize synthesis of the existing materials and, ultimately, guide the development of new ones. In addition, structure and properties of nanoclusters very often significantly differ from their bulk counterparts making them promising building blocks for the design of novel, cluster-assembled materials with tailored properties.[11] As an example, magnetic clusters such as iron(III) complexes with oxygen-based ligands (single molecule magnets)[12-15] have attracted considerable attention due to their potential applications in quantum computing[16] and spintronics.[17] The interaction between iron and oxygen is also one of the most important chemical processes in understanding corrosion, biological oxygen transport and several catalytic processes (see *e.g.* Refs 18-20). Consequently, structure, properties and reactivity of small iron oxide clusters and nanoparticles were subject of a number of experimental studies.[21-37] With only few exceptions for $(Fe_2O_3)_n$ with $n \leq 2$[37-39] computational investigations of structure and properties of iron oxide clusters were so far limited to non-stoichiometric ionic[40-44] and neutral[45,46] ones. Larger, neutral and stoichiometric $(Fe_2O_3)_n$ clusters with $n = 2-6$ and 10 were studied at the density functional theory (DFT) level considering only manually constructed and locally optimized structures.[47] However, atomic level characterization of larger clusters is complicated by a large number of possible structural isomers making manual construction of all possible structural models followed by local structure optimization a very challenging task.[10,48] Therefore, several techniques for automatic determination of the most stable cluster structures have been proposed that rely on global energy minimization strategies.[49,50] One of the most often used approaches is genetic algorithm (GA) that finds the global minimum structure by an evolutionary process.[10,50,51] GA combined with DFT has been applied successfully for structure predictions of various gas phase clusters (see *e.g.*, Refs 48 and 52). However, due to a challenging electronic structure and a very large number of possible spin configurations global structure optimizations of magnetic materials such as $(Fe_2O_3)_n$ clusters are computationally very demanding and to our best knowledge have not yet been reported for metal oxides.

In this work accurate, *ab initio* derived interatomic potential functions (IP) of Born-Mayer type[53] tailored for the description of $Fe_2O_3$ were used in combination with GA[48] for an extensive search through configuration space of $(Fe_2O_3)_n$ with $n$ = 1-5. A detailed description of the new IP, including derivation procedure and performance tests for the description of larger clusters, nanoparticles and bulk $Fe_2O_3$ polymorphs will be presented in a forthcoming publication. The global minimum structures obtained at the IP level were subsequently refined at the DFT level. All DFT calculations were performed using the TURBOMOLE program package[54-56] along with the B3-LYP exchange correlation functional.[57-59] The multipole accelerated resolution of identity (MARI-J) method[60] for the Coulomb term employing triple-zeta valence plus polarization (TZVP) basis sets for all atoms[61] was used along with the corresponding auxiliary basis sets.[62] Up to 50 most stable structures located by GA employing IP were first locally optimized assuming ferromagnetic (FM) states. Next, several most stable isomers were used for single point calculations of different spin states according to the broken symmetry approach of Noodleman *et. al.*[63-65] The construction of initial orbitals is the most crucial part in this approach.[66] Therefore, start orbitals were generated from FM states by first applying a Boys and Foster localization procedure on the valence (spin) orbitals,[67] assigning them to atoms and then flipping the spin at the desired Fe atoms.[68] The lowest energy magnetic (spin) configurations were subsequently determined employing the Ising-type Hamiltonian

$$H = -\sum_{i<j} J_{ij} S_i S_j. \qquad (1)$$

The exchange coupling constants $J_{ij}$ between Fe atoms $i$ and $j$ were obtained from least square fits of DFT energy differences between ferromagnetic and several broken symmetry spin states for each cluster size (see supporting information). For $(Fe_2O_3)_n$ clusters with $n$ = 2-5 the spins $S_i$ of $Fe^{3+}$ ions in Eq. 1 were set to ±5/2, whereas for $n$ = 1 also lower spin states of Fe atoms were taken into account. Similar approach has been applied to determine coupling constants of polynuclear transition metal clusters and complexes.[69-72] The computed spin expectation values $<S^2>$ are in good agreement with predicted values (see supporting information) indicating a weak overlap of magnetic orbitals, which supports the applied approach for the calculation of magnetic coupling constants.[72] The structures of the most stable spin configurations were locally optimized leading to the magnetic ground state (GS) for a cluster of a particular size. The binding energies $\Delta E_b$ of the clusters with respect to the GS of $Fe_2O_3$ were calculated as energy of the following reaction:

$$(Fe_2O_3)_n \rightarrow n\,Fe_2O_3 \qquad (2)$$

Figure 1 shows the most stable structures of $(Fe_2O_3)_n$ clusters along with isosurfaces of their GS spin density. Table I summarizes the relative energies of the GS and FM states of the clusters along with average magnetic coupling constants $\bar{J}$ (cf. Eq. 1). In order to analyze structural dependence of magnetic coupling constants Fe-Fe pairs in the clusters were divided into ones bridged by one and two O atoms, respectively. The mean Fe-O distance ($r_{Fe-O}$) and the difference of Fe-O-Fe bonding angles within twofold Fe-O ring ($\alpha_1$-$\alpha_2$) were used as structure parameters for Fe-Fe pairs bridged by one and two O atoms, respectively. Figure 2 shows the dependence of the calculated coupling constants $J_{ij}$ on the two structure parameters for $(Fe_2O_3)_n$ clusters with $n$ = 3-5. Generally, calculated coupling constants indicate predominantly antiferromagnetic (AF) and some weak FM interactions. The latter occur exclusively in twofold rings with both Fe-O-Fe angles $\alpha_1$ and $\alpha_2$ close to 90° (cf. Fig. 2). With an increasing difference $\alpha_1$-$\alpha_2$ the coupling constants become increasingly AF (see supporting information). All Fe-Fe pairs bridged by one O atom exhibit strong AF interactions that become weaker with increasing Fe-O distance.

For $Fe_2O_3$ the most stable configuration is the $C_{2v}$ symmetric, planar **1A** corresponding to the $^1B_1$ state. This structure was proposed as the most stable isomer of neutral[38,39] and anionic[35] $Fe_2O_3$ in previous studies along with an AF coupling of iron atoms. Natural population analysis[73] (see supporting information) reveal that **1A** formally consists of one $Fe^{3+}$ ($S$ = -5/2) bound via two $O^{2-}$ to $Fe^{2+}$ ($S$ = 4/2) and to the singly coordinated, terminal $O^-$ ($S$ = 1/2). Structure optimization of the corresponding triplet state results in an angled isomer **1B** that also formally consist of $Fe^{3+}$ ($S$ = -3/2), $Fe^{2+}$ ($S$ = 4/2), two $O^{2-}$ and one singly coordinated $O^-$ ($S$ = 1/2). Such an angled structure was suggested as the most stable configuration of the quartet spin state of $Fe_2O_3^+$.[36] The ferromagnetic $^{11}B_1$ state of $Fe_2O_3$ shows a planar structure similar to **1A** and is significantly (96.3 kJ mol$^{-1}$ $Fe_2O_3$) higher in energy than the $^1B_1$ state, indicating a very strong magnetic coupling between iron centers. In contrast to $Fe_2O_3$ natural population analysis performed for larger $(Fe_2O_3)_n$ clusters with $n$ = 2-5 reveal exclusively $Fe^{3+}$ species with well localized spin densities. The twofold coordinated Fe and singly coordinated O atom is a unique structural motif of $Fe_2O_3$ compared to the remaining $(Fe_2O_3)_n$ clusters. The distinct electron configuration of $Fe_2O_3$ renders comparison of magnetic coupling constants with values obtained for larger clusters less meaningful.

The most stable configuration of $(Fe_2O_3)_2$ is a cage-like, $C_{2v}$ symmetric **2A** with an antiferromagnetic $^1A_2$ GS that resembles the adamantane structure. The open, sheet-like $C_2$ symmetric **2B** with a ferrimagnetic $^{11}B$ GS is the second lowest configuration. For both **2A** and **2B** the FM states, $^{21}A_2$ ($T_d$) and $^{21}B_g$ ($C_{2h}$), respectively, are higher in energy. The magnetic coupling constant of -132 ± 3 cm$^{-1}$ calculated for **2A** (assuming $T_d$ symmetry) indicates a strong antiferromagnetic interaction arising from a large energy difference between the GS and the FM state. For **2B** three magnetic coupling constants were determined, two small FM values 1 ± 1 and 5 ± 1 cm$^{-1}$ as well as one AF interaction of -11 ± 1 cm$^{-1}$. In contrast to our findings, other studies that performed less extensive search of possible spin states reported a ferrimagnetic state with $C_{3v}$ symmetry[37] and a FM $^{21}A_1$ ($T_d$) state[47] of **2A** as the most stable configuration of neutral $(Fe_2O_3)_2$. For anionic $(Fe_2O_3)_2^-$ an AF configuration was found as the most stable.[44] **2B** was suggested as the second most stable configuration in earlier studies.[37,47] However, in contrast to the ferromagnetic, $C_2$ symmetric $^{11}B$ GS found in the present work these studies reported a FM, $^{21}B_g$ state with $C_{2h}$ symmetry as the GS.

For (Fe$_2$O$_3$)$_3$ the most stable **3A** exhibits C$_1$ symmetry along with an AF ground state. The second most stable isomer **3B** with an open, sheet-like C$_s$ symmetric structure shows an AF $^1$A' GS. The smaller mean coupling constant of -38 ± 1 cm$^{-1}$ for **3B** compared to **3A** (-67 ± 1 cm$^{-1}$) is due to a smaller energy difference between its AF ($^1$A') and FM ($^{31}$B$_2$) states. **3A** shows only strong AF interactions with $J_{ij}$ between -130 and -19 cm$^{-1}$. In contrast, for **3B** four symmetry inequivalent coupling constants were calculated, three indicating AF coupling and one weak FM interaction.

(Fe$_2$O$_3$)$_4$ shows two compact, C$_1$ symmetric isomers **4A** and **4B** with relative energy difference of only 4 kJ mol$^{-1}$ Fe$_2$O$_3$. For both structures the AF state is the most stable spin configuration with magnetic coupling constants of similar magnitude, predominant AF interactions (up to -139 cm$^{-1}$) and some weak FM couplings (less than 25 cm$^{-1}$). Consequently, both **4A** and **4B** show similar mean coupling constants and similar energy separations between the GS and FM states.

The two most stable isomers **5A** and **5B** of (Fe$_2$O$_3$)$_5$ are C$_1$ symmetric. The tower-like **5A** shows AF ground state whereas the compact structure **5B** exhibits a ferrimagnetic $^{11}$A GS. The second most stable spin configuration of **5B** that is only 1 kJ mol$^{-1}$ Fe$_2$O$_3$ less stable than the GS is AF. Note, that such a small energy difference is below the accuracy of the DFT method used and certainly below the zero point vibrational energy level. **5A** and **5B** show similar energy differences between their GS and FM states. Consequently, as in case of (Fe$_2$O$_3$)$_4$ they show similar average magnetic coupling constants. The calculated coupling constants in **5A** indicate exclusively AF coupling between Fe atoms with $J_{ij}$ in the range of -147 to -11 cm$^{-1}$. **5B** shows predominantly AF couplings with $J_{ij}$ up to -150 cm$^{-1}$ and few weak FM interactions ($J_{ij}$ less than 9 cm$^{-1}$).

The broken symmetry DFT approach used in the present study was shown to yield magnetic coupling constants in a good agreement with experimental data, in particular in combination with the B3-LYP exchange-correlation functional and triple-zeta valence basis sets (see Ref. 72 and references therein). The calculated magnetic coupling constants in (Fe$_2$O$_3$)$_n$ clusters are similar to the values in iron(III) oxide complexes, between -90 and 12 cm$^{-1}$, calculated using DFT and including spin orbit effects.[12] In contrast, bulk hematite exhibits weaker antiferromagnetic (up to -21 cm$^{-1}$) and FM interactions (less than 4 cm$^{-1}$) as determined by neutron scattering[74] and DFT calculations.[75-77]

Figure 3 shows comparison of IP and DFT results for (Fe$_2$O$_3$)$_n$ clusters with $n$ = 2-5, including both GS and FM states: average Fe-O bond lengths and binding energies for the most stable structures as well as relative stability of the two lowest energy isomers. The structures of the clusters are very similar in all three cases, IP and DFT for GS as well as FM states, with average Fe-O bond lengths deviating by less than 0.05 Å. In IP and DFT FM structures the Fe-O bond lengths are up to 0.04 Å longer than in the GS structures. As a general trend the average bond lengths increase with increasing cluster size, from 1.82 Å in **2A** to 1.90 Å in **5A**, and this behavior is very well reproduced by the IP. The small structural differences between GS and FM states demonstrate that the influence of magnetic states on the structure of larger (Fe$_2$O$_3$)$_n$ clusters is small. This supports our approach for computation of magnetic coupling constants, *i.e.*, single point DFT calculations on structures optimized for FM states, which are virtually independent of such small structural changes.

Figure 3b shows the cluster size dependence of binding energies $\Delta E_b$ for the most stable cluster structures. The values of $\Delta E_b$ calculated for FM states show almost constant shift with respect to the GS, consistent with the relative stabilities shown in Table 1. The values of $\Delta E_b$ calculated using IP are close to results for the GS and also very well reproduce the monotonic increase of $\Delta E_b$ with the cluster size. The IP also very well reproduces the relative stability of the two most stable isomers in their GS, in particular for $n$ = 4 and 5 (Fig. 3c). Therefore, GA in combination with the IP is particularly well suited for global structure optimizations of larger (Fe$_2$O$_3$)$_n$ cluster and nanoparticles. The differences between relative stabilities of the clusters evaluated for their GS and FM states decrease with increasing cluster size indicating that the precise magnetization state of the clusters has not only small influence on the geometric structure but also on the relative stabilities of larger (Fe$_2$O$_3$)$_n$ clusters.

In conclusion, global minimum structures of neutral (Fe$_2$O$_3$)$_n$ clusters with $n$ = 1-5 were determined employing GA in combination with *ab initio* parameterized IP and subsequent refinement at the DFT level. Systematic investigations of magnetic configurations of the clusters using the broken symmetry approach reveal antiferromagnetic and ferrimagnetic ground states. Whereas $(Fe_2O_3)_n$ clusters with $n$ = 2-5 contain exclusively $Fe^{3+}$, $Fe_2O_3$ was found to be a special case formally containing both $Fe^{2+}$ and $Fe^{3+}$. Magnetic coupling constants obtained from least square fits of energy differences between FM and broken symmetry states revealed predominantly strong AF interactions, which exceed bulk values found in hematite. The precise magnetization state of the clusters has small influence on their geometric structure as demonstrated by small structural deviations between their GS and FM states. Starting from $n$ = 4 also the relative energies of different cluster isomers are weakly influenced by their magnetic configuration. Therefore, simulations of the structures of larger $(Fe_2O_3)_n$ clusters and nanoparticles can be performed irrespective of their magnetization state.

**Acknowledgments**

Authors gratefully acknowledge financial support from the Fonds der Chemischen Industrie and Turbomole GmbH.

**FIGURES**

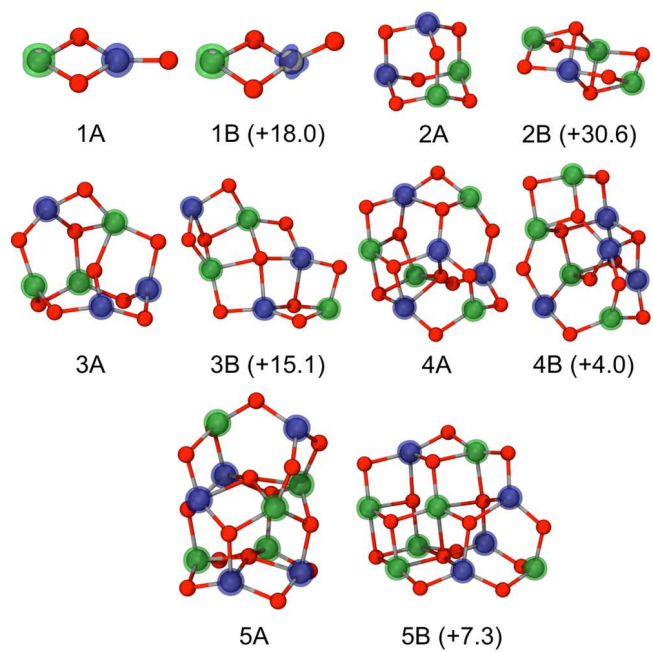

**Fig. 1** The most stable isomers of (Fe$_2$O$_3$)$_n$ clusters with $n$ = 1-5 along with the corresponding spin densities of the ground state (α-spin: green, β-spin: blue). Relative energies in parentheses (kJ mol$^{-1}$ Fe$_2$O$_3$). Fe: grey, O: red.

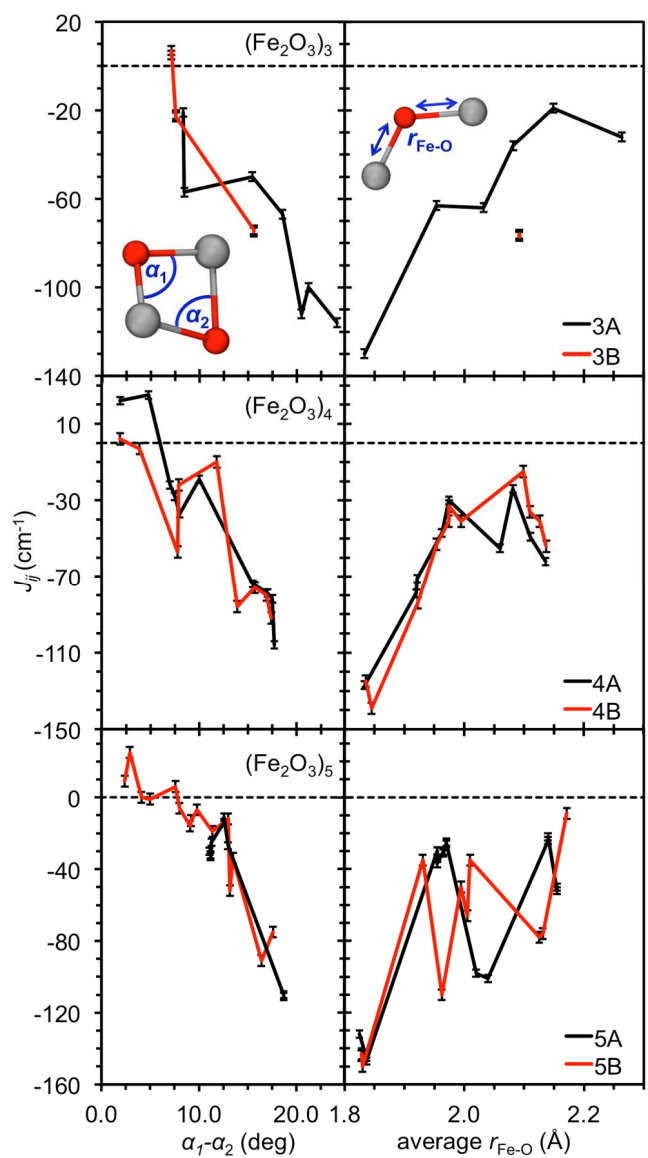

**Fig. 2** Magnetic coupling constants $J_{ij}$ for two most stable structures of $(Fe_2O_3)_n$ clusters with $n$=3-5. The coupling constants for Fe-Fe pairs bridged by two O atoms (left side) are shown as a function of the differences of Fe-O-Fe bond angles ($α_1$-$α_2$). The average Fe-O distance ($r_{Fe-O}$) is used as structure parameter for Fe-Fe pairs bridged by one O atom (right side). Standard deviations indicated by vertical bars. Fe: grey, O: red.

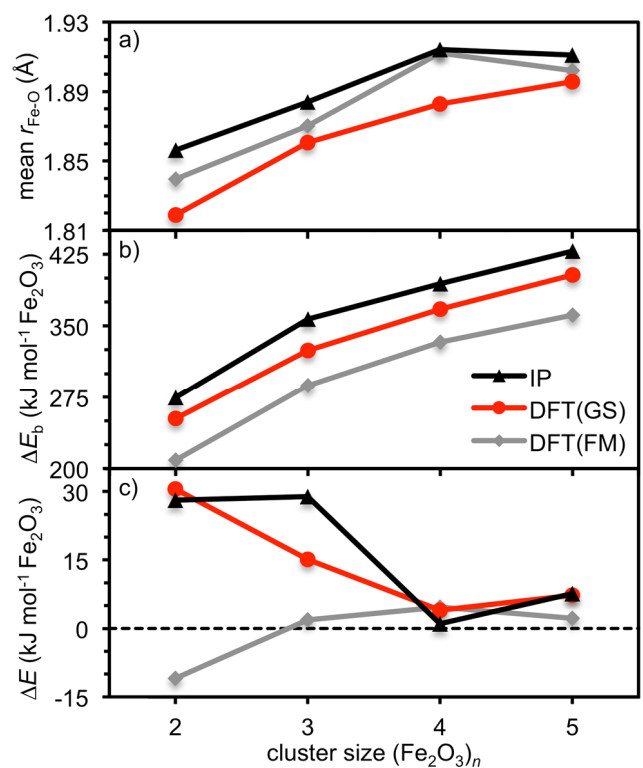

**Fig. 3** Comparison of IP and DFT results for $(Fe_2O_3)_n$ clusters with $n$ = 2-5: (a) average Fe-O bond lengths, $r_{Fe-O}$, (b) binding energies, $\Delta E_b$, and (c) relative stability, $\Delta E$, of the two lowest energy isomers ($n$A and $n$B, cf. Fig. 1). GS and FM indicate ground and ferromagnetic states, respectively.

**TABLES**

**Table 1** Relative energies $\Delta E$ (kJ mol$^{-1}$ Fe$_2$O$_3$) and average magnetic coupling constants $\bar{J}$ (cm$^{-1}$) for different structures and electronic states of (Fe$_2$O$_3$)$_n$ clusters.

| Structure | $\Delta E$ | State | $\bar{J}$ |
|---|---|---|---|
| 1A | 0.0 | $^1B_1$ (C$_{2v}$) | - |
| 1B | 18.0 | $^3A''$ (C$_s$) | |
|    | 96.3 | $^{11}B_1$ (C$_{2v}$) | |
| 2A | 0.0 | $^1A_2$ (C$_{2v}$) | -132 ± 3 |
|    | 43.6 | $^{21}A_2$ (T$_d$) | |
| 2B | 0.0 | $^{11}B$ (C$_2$) | -1 ± 1 |
|    | 2.1 | $^{21}B_g$ (C$_{2h}$) | |
| 3A | 0.0 | $^1A$ (C$_1$) | -67 ± 1 |
|    | 37.0 | $^{31}A$ (C$_1$) | |
| 3B | 0.0 | $^1A'$ (C$_s$) | -38 ± 1 |
|    | 23.9 | $^{31}B_2$ (C$_{2v}$) | |
| 4A | 0.0 | $^1A$ (C$_1$) | -47 ± 1 |
|    | 34.6 | $^{41}A$ (C$_1$) | |
| 4B | 0.0 | $^1A$ (C$_1$) | -54 ± 2 |
|    | 35.3 | $^{41}A$ (C$_1$) | |
| 5A | 0.0 | $^1A$ (C$_1$) | -53 ± 1 |
|    | 42.0 | $^{51}A$ (C$_s$) | |
| 5B | 0.0 | $^{11}A$ (C$_1$) | -45 ± 2 |
|    | 36.9 | $^{51}A$ (C$_1$) | |